\documentclass[12pt,epsfig]{article}

\pdfoutput=1
\usepackage{sitender}
\usepackage[margin=2cm]{geometry}

 \input epsf.sty
\title{The Mostly BRST exact Operator }
\author{}

\begin{document}

	\baselineskip 24pt
	
	\begin{center}
		{\Large \bf  {The Mostly BRST Exact Operator in 
			 Superstrings}}
		
	\end{center}
	
	\vskip .3cm
	\medskip
	
	\vspace*{.10ex}
	
	\baselineskip=18pt

	\begin{center}
		
		{Sitender Pratap Kashyap}
		
	\end{center}
	
	\vspace*{2.0ex}

\begin{center} 		\it The Institute of Mathematical Sciences\\
	IV Cross Street Road, CIT Campus, Taramani,
		Chennai 600113, India
\end{center}
	
	\vspace*{1.0ex}
	\centerline{ E-mail:  sitenderpk@imsc.res.in}
	
	\vspace*{5.0ex}

	\centerline{\bf Abstract}
	 \bigskip
	A careful gauge fixing of the conformal killing group (CKG) on genus zero surfaces in bosonic string theory gives non-vanishing two point amplitudes that match the corresponding field theory expressions \cite{Erbin:2019uiz,Seki:2019ycz}. An important ingredient for gauge fixing two point amplitudes in \cite{Seki:2019ycz} was the mostly BRST (mBRST) exact operator.  The utility of this operator in gauge fixing CKG perhaps is not just limited to two point amplitudes - we can insert a mBRST exact operator to fix a conformal killing vector (CKV) instead of fixing the position of a vertex operator in a general tree level bosonic string amplitude \cite{Seki:2021ivm,Kishimoto:2021csf}.
	
	Using the mBRST exact operator, written in the pure spinor variables, we get the expected two point superstring amplitudes \cite{Kashyap:2020tgx}. In this work we explore if it is possible to use this operator for fixing CKG for general tree level amplitudes in superstrings. We find it to be the case by explicitly re-calculating the three gluon amplitude in open strings by making use of this operator. Finally, we show that this operator directly emerges by following a Faddeev-Popov gauge fixing method in bosonic string theory. This derivation is independent of the number of external strings, thus proving the above claim. 
		\vfill

	%\begin{center}
	%{\Huge Preliminary version}
	%\end{center}
	
	\vfill \eject
	
	\baselineskip 18pt
	
	\tableofcontents

\section{Introduction}
One of the standard lores in the string perturbation theory, until very recently, was that tree level 2-point amplitudes are identically zero. However, it was realized in \cite{Erbin:2019uiz} that this understanding is incorrect - the string two-point tree level amplitudes are finite and indeed agree with the corresponding standard free particle amplitude in quantum field theories. Finiteness of the two point function is required for a few reasons - see \cite{Erbin:2021smf} for a nice summary. The scattering amplitude is generally written as 
\be 
S=I+iT
\ee
A consistent interpretation of the Polyakov prescription prior to \cite{Erbin:2019uiz} was that it represents the $iT$ part of the scattering amplitude. Finiteness of the two point amplitude means that the Polyakov prescription gives the "full scattering" amplitude $S$ and not just the $iT$ part. The authors first gave an intuitive argument why this must be the case and then they explicitly demonstrated this to be true by following Faddeev-Popov trick for open strings. Same result was obtained by employing operator methods \cite{Seki:2019ycz}. Both of these papers however talked about bosonic strings and assumed that the results will continue to hold in superstrings as well.

In \cite{Kashyap:2020tgx} it was shown that above conclusion holds true in even in superstrings by calculating two gluon/gluino amplitudes in the pure spinor formalism of superstrings\footnote{Lower point amplitudes on a disk in pure spinor formalism were calculated in \cite{Bischof:2020tnf}. In this paper one point closed string amplitude on a disk where the usual prescription \eqref{ampPS} is not applicable was calculated using an alternate prescription \cite{Berkovits:2016xnb}. It would be interesting to see how this approach is related to the one where mBRST operator is used}. The pure spinor formalism (PS) is the only super-Poincare covariant approach to quantize superstrings \cite{Berkovits:2000fe}. It has been argued many times that PS formalism is equivalent to the other two important formalisms of string theory namely the RNS and the Green-Schwarz. It proves advantageous to keep all the spacetime symmetries of the underlying theory manifest. This has allowed many non-trivial and difficult computations to be performed in the pure spinor formalism. One disadvantage however, until very recently has been that there was no gauge invariant action for pure spinor superstring. This problem seems to have been solved in \cite{Jusinskas:2019vmd}. It has been shown that this gauge invariant action does reproduce the standar pure spinor action upon suitable gauge fixing. However, a direct Polyakov like prescription to arrive at the pure spinor amplitudes is still not available. The tree amplitude in pure spinor formalism (open string) are computed via 
\be 
\mc{A}_{n\ge 3} =\Big\la V_1(z_1) V_2(z_2) V_3(z_3)\int dz_4 U(z_4)\cdots \int dz_n U(z_n)\Big\ra \label{ampPS}
\ee
where $V_i(z_i)$ are the unintegrated vertex operators and $U(z_i)$ are the integrated vertex operators - this has the same structure as in the bosonic string theory. However, this formula implicitly assumes that there are at least three vertex operators. A naive application of the above formula for two strings gives a vanishing answer. For consistency with other string theory formalisms this must not be the case. Because of absence of Polyakov like prescription, it is not clear how to calculate two point amplitudes. Thus, in order to make progress we need to come up with a prescription for calculating the two point function. This was achieved in \cite{Kashyap:2020tgx}. The basic idea is based on the simple observation that the pure spinor formalism can be reinterpreted as an $N=2$ topological string \cite{Berkovits:2005bt}. The amplitude prescription of $N=2$ topological theory is based on the bosonic string theory.  Consequently it seems natural to borrow the amplitude prescription of the two point function from the bosonic string. The crucial element that allows us to guess the correct prescription is the mostly BRST exact operator, first introduced in \cite{Seki:2019ycz}
  \be 
  \mc{V}_{0}(z)\equiv \int_{-\infty}^{\infty} \f{dq}{\pi \a' q} \;[Q_B,e^{-iqX^0(z)}] \label{mBRSTbosonic}
  \ee
where, $Q_B$ is the BRST charge in the bosonic string\footnote{We shall use the notation $[Q_{B},O(x)]\equiv \oint dy j_{B}(y)\, O(x)$, where $j_{B}$ is the BRST current.}. The integrand of this operator is BRST exact if $q\ne 0$ - this is why $\mc{V}_0$ is called mostly BRST exact operator.  Written in this form we can simply uplift this relation to the pure spinor superstring \cite{Kashyap:2020tgx} 
\be 
V_0(z)\equiv \f{1}{2\pi\a'}\int_{-\inf}^{\inf}dq (\l\g^0\t)e^{iq X^0}    \label{mBRSTPS}
\ee
where, the r.h.s of above equation is written after action of the BRST charge $Q=\oint dz \l^\a d_\a $ of the pure spinor formalism on $e^{iqX^0}$. In this paper we shall work with a covariant expression
\be 
V_0(z)=\f{\a'}{4\pi i}\int_{-\infty}^{\infty}dq\; t_m\rb{\l\g^m\t}:e^{-iq t_n X^n(z)}: \, ,  \qquad t^2=-1										\label{mBRSTcov}
\ee
 This differs from the \eqref{mBRSTPS} by a factor of $\a'^2/2i$  (see section \ref{mBRSTintro} for more details where this factor follows neatly).

The operator \eqref{mBRSTbosonic} has now been found to be of more general utility than just providing correct two point amplitude. Using this we can fix the CKG in a more general tree level amplitude. Let us explain what we mean by this. The amplitude \eqref{ampPS} can be evaluated by 
 \be 
\mc{S}_{n\ge 3} =\Big\la V_0(z_0) V_1(z_1) V_2(z_2)\int dz_3 U(z_3)\cdots \int dz_n U(z_n)\Big\ra  \label{genamp}
\ee
 where, $V_0$ is the operator in \eqref{mBRSTbosonic}. This was shown for 3 point tachyon amplitude and the Veneziano amplitude in \cite{Seki:2021ivm,Kishimoto:2021csf}. In this work we explore if similar conclusions hold in the superstrings by making use of \eqref{mBRSTcov} in the pure spinor formalism. In section \ref{gen_amp} we start by computing the three gluon amplitude in two ways - standard three point calculation using \eqref{ampPS} and  then using \eqref{genamp}. We find an agreement upto overall sign. In the rest of the paper we attempt to prove this for general $n$. We shall follow a method similar to the Faddeev-Popov method of \cite{Erbin:2019uiz} to gauge fix the two point amplitude, but, with a slight modification that gives rise to $c$ ghosts so that it is easy to identify the BRST exact operator. We shall see that the procedure is immune to how many other extra vertex operators are there - this proves the result for general $n$. This procedure works well for open strings. For closed strings this does not work straightforwardly. We identify the cause and propose a solution in section \ref{closed_strings}. In the discussion section \ref{discuss} we mention some further uses of this operator.

\section{Three gluon amplitudes using mBRST exact operator}  \label{gen_amp}
In this section we are going to do calculations involving gluons by making use of the pure spinor formalism.  Here we shall not introduce PS formalism. We shall only recall the details that we require in appendices \ref{ps_results} and \ref{3pt} . See \cite{Berkovits_ictp,Joost_thesis,Mafra:2009wq,Oliver_thesis} for detailed introduction to the pure spinor formalism. First we are going to calculate the three gluon amplitude via the standard prescription. In the next subsection, we calculate the generalized amplitude using three gluon vertex operator and one mBRST exact operator. We find these agree with each-other upto a sign.  

The standard prescription to calculate the three gluon amplitude is to evaluate the correlation function of three gluon vertex operators fixed at three points on the boundary of a disk. 
\be 
\mc{A}_{3}=\la V_1(z_1) \, V_2(z_2) \, V_3(z_3)\ra + (2\leftrightarrow 3)  \label{3pt_total}
\ee
where, $V=\l^\a A_\a(\t)e^{ik\cdot X}$ are the unintegrated vertex operators, $z_i$ are coordinates along the real line and $A_{\a}$ is the spinor superfield appearing in covariant 10 dimensional super-Yang Mills (sYM) equation of motion (e.o.m). The coordinates $z_1,z_2$ and $z_3$ can be arbitrarily chosen - this fixes the conformal killing group on a disk. For convenience we choose $z_1=0,z_2=1,z_3=\infty$. On the r.h.s of \eqref{3pt_total} the two terms correspond to the two ways in which three vertex operators can be placed at the boundary of a disk.  We present the details of the computation in appendix {\ref{3pt}}\footnote{Even though its a standard known result, we recalculate it for uniformity in conventions}.  The result for one of the orders is 
\be 
\mc{A}^{I}_{BBB}(\e_1,k_1; \e_2,k_2; \e_3,k_3)
=\f{i }{180}\Big[ \rb{\e_1\cdot \e_2}\rb{\e_3\cdot k_{1}}+\rb{\e_1\cdot \e_3}\rb{\e_2\cdot k_{3}}+\rb{\e_2\cdot \e_3}\rb{\e_1\cdot k_{2}}\Big] \ug_3\label{3gluon_amp}
\ee
where, $\e_i$ and $k_i$ are the polarization tensor and the momentum for the $i^{\textup{th}}$ gluon respectively. $\ug_3$ is proportional to the momentum conserving delta function \eqref{g3}. 

\subsection{Three point amplitude using mBRST exact operator}
In this subsection we shall make use of the mBRST exact operator to calculate the 3 gluon amplitude
\be 
\mc{S}_{4}&\equiv&\int_{-\inf}^{\inf} dz_3 \; \la V_0(z_0) V_1(z_1) V_2(z_2) U(z_3)\ra
\ee
The unintegrated and integrated vertex operators are given by 
\be 
V\equiv\l^\a A_\a,\quad U=\f{4}{\a'}\rb{\f{1}{2}\p\t^\a A^{3}_\a+\Pi^m A^{3}_m -d_\s W^{\s}_3 +N^{ mn}\mc{F}^3_{mn}}
\ee
where, $A_\a, A_m, W^\a$ and $\mc{F}_{mn}\equiv \p_m A_n- \p_n A_m$ are the superfields appearing in the super Yang Mills equation of motion. Consequently 
\be 
\mc{S}_{4}&=&\f{1}{i\pi}\int_{-\inf}^{\inf}dq \int_{-\inf}^{\inf} dz_3 \biggl\la \left[\rb{\l \g^m\t}t_m \,e^{iqX^n t_n }\right](z_0)\; \rb{\l^\a A^{1}_{\a}}(z_1) \; \rb{\l^\b A^{2}_{\b}}(z_2) \non\\
&&\times\rb{\f{1}{2}\p\t^\a A^{3}_\a+\Pi^m A^{3}_m -d_\s W^{\s}_3 +N^{ mn}\mc{F}^3_{mn}}(z_3)\biggl\ra \label{s4}
\ee
 We shall drop the arguments $z_0,z_1,z_2,z_3$ from now on to avoid cluttering of notation. 
In order to evaluate the above, we need to calculate the following CFT correlators
\be 
T_1&\equiv& \biggl\la \left[\rb{\l \g^m\t}t_m \,e^{iqX^n t_n }\right]\; \rb{\l^\a A^{1}_{\a}} \; \rb{\l^\b A^{2}_{\b}} \p\t^\a A^{3}_\a \biggl\ra \label{T1}\\
T_2&\equiv& \biggl\la \left[\rb{\l \g^m\t}t_m \,e^{iqX^n t_n }\right]\; \rb{\l^\a A^{1}_{\a}} \; \rb{\l^\b A^{2}_{\b}} \Pi^m A^{3}_m \biggl\ra \label{T2} \\
T_3&\equiv& -\biggl\la \left[\rb{\l \g^m\t}t_m \,e^{iqX^n t_n }\right]\; \rb{\l^\a A^{1}_{\a}} \; \rb{\l^\b A^{2}_{\b}} \;d_\s W^{\s}_3  \biggl\ra \label{T3}\\
T_4&\equiv& \biggl\la \left[\rb{\l \g^m\t}t_m \,e^{iqX^n t_n }\right]\; \rb{\l^\a A^{1}_{\a}} \; \rb{\l^\b A^{2}_{\b}} \;N^{mn}\mc{F}^3_{mn}  \biggl\ra \label{T4} 
\ee
$T_1$  gives vanishing contribution as there are no $p_\a$ to eliminate $\p \t^\a$. $T_2,T_3$ and $T_4$ are non-vanishing and we shall calculate these one by one. It is convenient to define   
\be 
\rb{\l \g^m\t}t_m \,e^{iqX^n t_n }\equiv \l^\a A^{\rb{0}}_\a,\quad \; A^{(0)}_{\a}=t_m\rb{\g^m\t}_\a e^{iqX^n t_n}\, \quad k_0^m\equiv q t^m   
\ee
 where $t^m$is a time like vector\footnote{Due to this $k_0$ is like a momentum which supplies a conformal weight $\a' k_0^2=-\a'q^2$ to $V_0$.}. Each of the factors in each of the terms in $T_2,T_3$ and $T_4$ have an accompanying factor $e^{ik_{j}.X}$ which can be evaluated to give \cite{Polchinski:1998rq}
\be
\left\la \prod_{j=0}^4 :e^{i k_j.X(z_j)}:\right\ra = i C_{D_2} \; (2\pi)^d \;\d(\sum_{j}k_j)\; \prod_{i<j} |z_i -z_j|^{2\a' k_i\cdot k_j} \equiv \Gamma_4
\ee
where, $k_i^2=0$ for $i=1,2,3$ represent the momenta of the gluons and $:\;:$ stands for boundary normal order. It will be convenient for us to define Mandelstam variables
\be 
s&\equiv& -(k_0+k_1)^2=  q^2 -2k_0\cdot k_1=-2k_2\cdot k_3 \non\\
t&\equiv& -(k_0+k_2)^2=  q^2 -2k_0\cdot k_2= -2k_1\cdot k_3 \label{mandelstam1}\\
u&\equiv& -(k_0+k_3)^2=  q^2 -2k_0\cdot k_3=-2k_1\cdot k_2 \non
\ee
where, the last equality in each of the lines above follows from momentum conservation and $k_1^2=k_2^2=k_3^2=0$. These add up to $s+t+u=3 q^2 -2k_0\cdot \rb{k_1+k_2+k_3}=3 q^2 +2k_0^2=q^2$. Let us begin by evaluating $T_2$
\be 
T_2&=& \biggl\la \rb{\l^\a A^{0}_\a}\; \rb{\l^\a A^{1}_{\a}} \; \rb{\l^\b A^{2}_{\b}} :\Pi^m A^{3}_{m}:\biggl\ra \non\\
&=&i\a'\sum_{j=0}^2\f{k^{m}_j}{z_j-z_3}\left\la   \rb{\l^\a A^{0}_\a}\; \rb{\l^\a A^{1}_{\a}} \; \rb{\l^\b A^{2}_{\b}} A^{3}_{m} \right\ra \ug_4\non\\
&\equiv &i\a'\sum_{j=0}^2\f{k^{m}_j}{z_j-z_3}\left\la  V_0 \; V_1\; V_2\; A^{\rb{3}}_{m} \right\ra \ug_4  \label{T2PSS}
\ee
where, $V_i\equiv \rb{\l^\a A^i_\a}$ and we used the OPE \eqref{piOPE} and used the generalized normal ordering defined as 
\be  
:AB:(x)\equiv \f{1}{2\pi i}\oint_{x}\f{dy}{y-x}A(y)B(x)
\ee
for coincident operators $A$ and $B$. Repeating similar steps and on making use of the OPE in \eqref{dOPE}, we find
\be 
T_3&=&-\f{\a'}{2} \left\la \f{D_\s V_0(z_0)}{z_0-z_3} V_1(z_1) V_2(z_2) W^{(3)\s}(z_3)\right\ra+\f{\a'}{2} \left\la V_0(z_0) \f{D_\s V_1(z_1)}{z_1-z_3}  V_2(z_2) \; W^{(3)\s}(z_3)\right\ra \non\\
&&-\f{\a'}{2} \left\la V_0(z_0)\;V_1(z_1)\; \f{D_\s V_2(z_2)}{z_2-z_3} \;  W^{(3)\s}(z_3)\right\ra \non\\
&=&-\f{\a' t_s}{2z_{03}} \left\la V_1(z_1) V_2(z_2) \rb{\l\g^s W^{(3)}}(z_3)\right\ra+\f{\a'}{2z_{13}} \left\la V_0(z_0) D_\s V_1(z_1)  V_2(z_2) \; W^{(3)\s}(z_3)\right\ra \non\\
&&-\f{\a'}{2z_{23}} \left\la V_0(z_0)\;V_1(z_1)\; D_\s V_2(z_2)\;  W^{(3)\s}(z_3)\right\ra \equiv T_{3a}+T_{3b}+T_{3c}                   \label{T3total}
\ee
where, we used $D_\s V_0=D_\s\big[(\l \g^s \t) t_s\big]=\l^\a (\g^s)_{\a\b} D_\s \t^\b \;t_s=(\l\g^s)_\s t_s$. Also, $T_{3a},T_{3b}$ and $T_{3c}$ correspond to the respective terms in the previous equality. $T_{3b}$ and $T_{3c}$ can be further simplified, but, we shall do it after presenting the result for $T_4$. This time on making use of the OPE in\eqref{NOPE}
\be 
T_4&=& \biggl\la V_0 \; V_1 \; V_2  \; :N^{mn}F^{(3)}_{mn}: \biggl\ra \non\\
&=&  \f{\a' \;t_s}{4z_{03}}\biggl\la (\l\g^{m n} \g^s \t)\; V_1  \; V_2  \; F^{(3)}_{mn}  \biggl\ra + \f{\a' }{4z_{13}}\biggl\la  V_0 \; (\l\g^{m n} A^{1})\; \; V_2 \; F^{3}_{mn}  \biggl\ra\non\\
&&+\;\; \f{\a' }{4z_{23}}\big\la  V_0 \; V_1 \;(\l\g^{m n} A^{3}) \; F^{3}_{mn}  \big\ra \equiv T_{4a}+T_{4b}+T_{4c} \label{T4t}
\ee 
Things can be simplified a further by noticing that the last two terms in $T_4$ cancel with last two terms of $T_3$. In order to see this we make use of the sYM e.o.m  
\be 
D_\a A_\b + D_{\b} A_\a =2(\g^m)_{\a\b}A_m \implies D_\a\rb{\l A^{(j)}}=D_\a\rb{V_j}= -\rb{\l D}A^{(j)}_\a +2\rb{\l\g^m}_{\a}A^{(j)}_m
\ee
where, $(\l D)$ represents the action of the integrand of  BRST charge
\be 
Q=\oint dz \l^\a d_\a
\ee
With this, 
\be 
T_{3b}&=&\f{\a'}{2z_{13}} \left\la V_0  \sqb{-\rb{\l D}A^{1}_\s +2\rb{\l\g^m}_{\a}A^{1}_m}  V_2 \; W^{\s}_3\right\ra \non\\
&=&-\f{\a'}{2z_{13}} \left\la V_0 \sqb{\rb{\l D}A^{1}_\s } V_2\; W^{\s}_3\right\ra +\f{\a'}{z_{13}} \left\la V_0 \sqb{\rb{\l\g^m}_{\s}A^{1}_m}  V_2 \; W^{\s}_3\right\ra  \non\\
&=&\f{\a'}{2z_{13}} \left\la  V_0 A^{1}_\s  V_2 \; \sqb{\rb{\l^\a D_\a} W^{\s}_3}\right\ra +\f{\a'}{z_{13}} \left\la V_0 A^{1}_m  V_2 \rb{\l\g^m W_{3}} \; \right\ra  \non\\
&=&-\f{\a'}{4z_{13}} \left\la  V_0 A^{1}_\s  V_2 \; \sqb{\l^\a \rb{\g^{st}}_\a^{\;\;\s}F^{3}_{st}}\right\ra +\f{\a'}{z_{13}} \left\la V_0 A^{1}_m  V_2 \rb{\l\g^m W_{3}} \; \right\ra  \non\\
&=&-\f{\a'}{4z_{13}} \left\la  V_0  \rb{\l\g^{st}A^{1}}  V_2 \; F^{3}_{st}\right\ra +\f{\a'}{z_{13}} \left\la V_0  A^{1}_m  V_2 \rb{\l\g^m W_{3}} \; \right\ra  \label{T3bs}
\ee
where, we used that $\rb{\l D}V_{j}=0, \; j=0,1,2$. For $j=1,2$ this is BRST closedness of gluon vertex operators, while for $V_0$ it follows from pure spinor constraint $\l\g^m\l=0$. Similarly,
\be 
T_{3c}&=&-\f{\a'}{2z_{23}} \left\la V_0  \;  V_1\sqb{-\rb{\l D}A^{2}_\s +2\rb{\l\g^m}_{\a}A^{2}_m}\; W^{\s}_{3}\right\ra \non\\
&=&-\f{\a'}{4z_{23}} \left\la  V_0 \;  V_1 \; \rb{\l\g^{st}A^{2}}F^{3}_{st}\right\ra -\f{\a'}{z_{23}} \left\la V_0 \;   V_1 \;A^{2}_m\; \rb{\l\g^m W_{3}} \; \right\ra   \label{T3cs}
\ee
Hence, on substituting \eqref{T3bs} and \eqref{T3cs} into \eqref{T3total}
\be 
T_3&=&-\f{\a't_s}{2z_{03}} \left\la V_1  V_2 \rb{\l\g^s W_{3}}\right\ra
-\f{\a'}{4z_{13}} \left\la  V_0 \rb{\l\g^{st}A^{1}}  V_2 \; F^{3}_{st}\right\ra +\f{\a'}{z_{13}} \left\la V_0 A^{1}_m  V_2 \rb{\l\g^m W_{3}} \; \right\ra \non\\
&& -\f{\a'}{4z_{23}} \left\la  V_0 \;  V_1 \; \rb{\l\g^{st}A^{2}}F^{3}_{st}\right\ra -\f{\a'}{z_{23}} \left\la V_0 \;   V_1 \;A^{2}_m\; \rb{\l\g^m W_{3}} \; \right\ra \label{T3totals}
\ee
Thus, on adding all the terms \eqref{T2PSS}, \eqref{T3totals} and \eqref{T4t} we find 
\be 
&&T_1+T_2+T_3+T_4\non\\
&=&\f{\a'}{z_{03}}\biggl[i k_0^m t_s\left\la  \rb{\l\g^s\t} \; V_1\; V_2\; A^{3}_{m} \right\ra - \f{t_s}{2} \left\la V_1 V_2 \rb{\l\g^s W_{3}}\right\ra \non\\
&&+\f{t_n}{2} \Big\la (\l \g^m \t)\; V_1 \; V_2 \; i\rb{k_{3m} A^{3}_{n}-k_{3n} A^{3}_{m}}\Big\ra +\f{t_s}{2} \Big\la (\l\g^{m n s} \t)\; V_1 \; V_2 \; ik_{3m}A^{3}_{n}  \Big\ra\biggl] \non\\
&&+\f{\a't_s}{z_{13}} \Big[i k_1^m \left\la  \rb{\l\g^s\t} \; V_1\; V_2\; A^{3}_{m} \right\ra+\left\la \rb{\l\g^s\t} A^{1}_m  V_2 \rb{\l\g^m W_{3}} \; \right\ra\Big]\non\\
&&+\f{\a'}{z_{23}} \Big[i k_2^m t_s\left\la  \rb{\l\g^s\t} \; V_1\; V_2\; A^{3}_{m} \right\ra - \left\la V_0 \;   V_1 \;A^{(2)}_m\; \rb{\l\g^m W^{(3)}} \; \right\ra\Big] \label{T1234PSS}
\ee
This expression can now be simplified by picking up the $\rb{\l^3 \t^5}$ components of the superfields and making use of the pure spinor superspace identities \eqref{id1} and \eqref{id2}. This computation by hand is tedious, but, fortunately can be handed over to computer algebra systems. We make use of CADABRA\cite{Peeters1,Peeters2}.  The result is (see appendix \ref{3pt_BRST exact} for details)
\be 
&&\hspace{-.4in}T_1+T_2+T_3+T_4\non\\ [.1in]
&=&\f{\a' }{360}\Big[ \rb{\e_1\cdot \e_2}\rb{\e_3\cdot k_{12}}+\rb{\e_1\cdot \e_3}\rb{\e_2\cdot k_{31}}+\rb{\e_2\cdot \e_3}\rb{\e_1\cdot k_{23}}\Big]\times \rb{\f{\rb{k_3\cdot t} }{z_{03}}+\f{\rb{k_2\cdot t} }{z_{13}}+\f{\rb{k_1\cdot t} }{z_{23}}}\non\\ [.1in]
&\equiv& -i \a'\mc{M}^{I}_{BBB}(\e_1,k_1; \e_2,k_2; \e_3,k_3) \times \rb{\f{\rb{k_3\cdot t} }{z_{03}}+\f{\rb{k_2\cdot t} }{z_{13}}+\f{\rb{k_1\cdot t} }{z_{23}}}
\ee
where $k_{ij}\equiv k_{i}-k_{j}$ and 
\be 
\mc{M}^{I}_{BBB}(\e_1,k_1; \e_2,k_2; \e_3,k_3)=\f{i }{360}\Big[ \rb{\e_1\cdot \e_2}\rb{\e_3\cdot k_{12}}+\rb{\e_1\cdot \e_3}\rb{\e_2\cdot k_{31}}+\rb{\e_2\cdot \e_3}\rb{\e_1\cdot k_{23}}\Big]
\ee
Consequently the full amplitude is given by 
\be
\mc{S}_{4}
&=& -\f{\a'}{\pi} \; \mc{M}^{I}_{BBB}\int_{-\inf}^{\inf}dq \int_{-\inf}^{\inf} dz_3 \rb{\f{\rb{k_3\cdot t} }{z_{03}}+\f{\rb{k_2\cdot t} }{z_{13}}+\f{\rb{k_1\cdot t} }{z_{23}}}\non\\
&&\times i C_{D_2} (2\pi)^{10}\,\d(\sum_{j}k_j) \prod_{i<j} |z_i -z_j|^{2\a' k_i\cdot k_j}
\ee
In order to proceed further we make the choice $z_0=0, z_1=1, z_2=\infty$. With this  
$$
\prod_{i<j} |z_i -z_j|^{2\a' k_i\cdot k_j} = |z_3|^{\a'\rb{s+t}}|1-z_3|^{-\a' t}
$$
Thus, we have 
\be
\mc{S}_{4}&=& -\rb{i C_{D_2} \; (2\pi)^{10} \mc{M}^{I}_{BBB}}\f{\a'}{\pi}\int_{-\inf}^{\inf}dq \d(\sum_{j}k_j)\int_{-\inf}^{\inf} dz_3 \rb{-\f{\rb{k_3\cdot t} }{z_{3}}+\f{\rb{k_2\cdot t} }{1-z_{3}}} |z_3|^{\a'\rb{s+t}}|1-z_3|^{-\a' t} \non\\
&\equiv&- \rb{i C_{D_2} \; (2\pi)^{10} \mc{M}^{I}_{BBB}}\f{\a'}{\pi}\int_{-\inf}^{\inf}dq \d(\sum_{j}k_j)\times \mc{I}(q,s,t,u)
\ee
where, we defined
\be 
\mc{I}(q,s,t,u)&\equiv&\int_{-\inf}^{\inf} dy \rb{-\f{\rb{k_3\cdot t} }{z_{3}}+\f{\rb{k_2\cdot t} }{1-z_{3}}} |y|^{\a'\rb{s+t}}|1-y|^{-\a' t}\non\\
&=&\rb{k_3\cdot t}\,B(\a'(s+t),-\a's)+\rb{k_2\cdot t} \,B(1+\a'(s+t),-\a's)\non\\
&&-\rb{k_3\cdot t}\,B(\a'(s+t),1-\a't)+\rb{k_2\cdot t}\,B(1+\a'(s+t),-\a't)\non\\
&&-\rb{k_3\cdot t}\,B(-\a's,1-\a't)-\rb{k_2\cdot t}B(-\a's, -\a't)
\ee
In the above equation $B(a,b)$ is the Euler Beta function defined as
\be
B(a,b)=\int_0^1 dx \,x^{a-1}(1-x)^{b-1}
\ee
We can simplify this further by noting that $$\rb{k_3\cdot t}=-\f{\rb{k_3\cdot k_0}}{q}=\f{-q^2+u}{2q}=-\f{q}{2}+\f{u}{2q},\quad \rb{k_2\cdot t}=-\f{\rb{k_2\cdot k_0}}{q}=\f{-q^2+t}{2q}=-\f{q}{2}+\f{t}{2q}$$
The term $\propto q$ in the above expression when integrated with $\rb{q+k_1^0+k_2^0+k_3^0}$ vanish due to the requirement $\sum_{i=1}^3k_i^0=0$
Thus, we have 
\be 
\mc{I}(q,s,t,u)
&=&\f{1}{2\a'q}\Big[-\a'\rb{s+t} \,\,B(\a'(s+t),-\a's)+\a't  \,B(1+\a'(s+t),-\a's)\non\\
&&+\a' (s+t) \,B(\a'(s+t),1-\a't)+\a't B(1+\a'(s+t),-\a't) \non\\
&&+\a' \rb{s+t} \,B(-\a's,1-\a't)-\a' t\,B(-\a's, -\a't)\Big]        \label{I3}
\ee
It can be seen that each line in the square bracket of \eqref{I3} is identically zero by making use of the following property of the Beta function 
\be 
B(x+1,y)=\f{x}{x+y}\,B(x,y),\quad B(x,y+1)=\f{y}{x+y}\,B(x,y)
\ee
This however does not mean that the total amplitude vanishes. We need to extract the $q=0$ contribution\footnote{For $q\ne 0$ this is essentially a statement of BRST of the operator.}. In order to get the physical result we need to examine the behavior of the amplitude near $q=0$. We simply expand $B(a,b)$ in rational fractions and then make use of the following formula
\be 
\f{1}{x-i\e}=\mc{P}\rb{\f{1}{x}} +\pi i \,\d(x)
\ee
where, $\mc{P}$ denotes the principal value of the argument. In our case we find  
\be 
\mc{I}&=&\f{\pi i}{2\a'^2}\d(q)\sqb{\rb{\f{k_3^0}{|k_3^0|}+ \f{k_3^0}{|k_1^0|}} +\rb{\f{k_2^0}{|k_1^0|}}- \rb{\f{k_3^0}{|k_3^0|}} +\rb{\f{k_2^0}{|k_2^0|}}-\rb{\f{k_3^0}{|k_1^0|}}-\rb{\f{k_2^0}{|k_1^0|}+ \f{k_2^0}{|k_2^0|}}+\cdots}\non\\
&=&0+\cdots\label{Ifinal}
\ee
where, $\cdots$ represent the polynomial in $q$.
This once again implies that the total generalized amplitude $\mc{S}_4$ is zero. But, if we look at it carefully this is what we must get. The gluon amplitude \eqref{3gluon_amp} is anti-symmetric in $2\leftrightarrow 3$ for the Abelian group. \eqref{3gluon_amp} gives the order $1,2,3$, but, we must include the order $1,3,2$ to this to get the total amplitude. 
\be 
&&\hspace{-0.4in}\mc{A}^{II}_{BBB}(\e_1,k_1; \e_3,k_3; \e_2,k_2)\non\\
&=&
\f{i }{180}\Big[ \rb{\e_1\cdot \e_3}\rb{\e_2\cdot k_{1}}+\rb{\e_1\cdot \e_2}\rb{\e_3\cdot k_{2}}+\rb{\e_3\cdot \e_2}\rb{\e_1\cdot k_{3}}\Big] \ug_3\non\\
&=&-\f{i }{180}\Big[ \rb{\e_1\cdot \e_3}\rb{\e_2\cdot k_{3}}+\rb{\e_1\cdot \e_2}\rb{\e_3\cdot k_{1}}+\rb{\e_2\cdot \e_3}\rb{\e_1\cdot k_{2}}\Big]\non\\
&=&-\mc{A}^{I}_{BBB}(\e_1,k_1; \e_2,k_2;\e_3,k_3)
\ee
This means that the total gluon amplitude vanishes indeed. Because the position of the 3rd gluon operator in \eqref{s4} was integrated from $-\infty$ to $\infty$, the amplitude we have constructed naturally takes care of both the orders in which vertex operators are to be placed on the boundary of the disk. In order to further confirm that our results are in perfect agreement with three point ordered amplitude, we note that the first 4 round bracketed terms inside the square bracket in \eqref{Ifinal} which arise from $-\infty<z_3<1$ give the order $1,2,3$  while the last two terms gives the contribution from $1<z_3<\infty$ i.e. the order $1,3,2$. Explicitly separating the contributions of these two orders we have
\be 
\mc{I}_{123}=\f{\pi i}{2\a'^2}\d(q)\sqb{\f{k_2^0}{|k_2^0|}- \f{k_1^0}{|k_1^0|}} 
\ee 
while, 
\be 
\mc{I}_{132}=\f{\pi i}{2\a'^2}\d(q)\sqb{\f{k_1^0}{|k_1^0|}- \f{k_2^0}{|k_2^0|}} 
\ee 
Hence, we conclude that 
\be
\mc{S}^I_{4}&=&-\rb{i C_{D_2} \; (2\pi)^{10} \mc{M}^{I}_{BBB}}\f{\a'}{\pi}\int_{-\inf}^{\inf}dq \d(\sum_{j}k_j) \times \f{\pi i}{2\a'^2}\d(q)\sqb{\f{k_2^0}{|k_2^0|}- \f{k_1^0}{|k_1^0|}}\non\\
&=&-\rb{i C_{D_2} \; (2\pi)^{10} \d(\sum_{i=1}^3k_i)\mc{M}^{I}_{BBB}}\times \f{i}{2\a'} \sqb{\f{k_2^0}{|k_2^0|}- \f{k_1^0}{|k_1^0|}}\non\\
&=&\f{i}{2\a'} \sqb{\f{k_1^0}{|k_1^0|}- \f{k_2^0}{|k_2^0|}}\times \mc{A}^{I}_{BBB}
\ee
and 
\be 
\mc{S}^{II}_{4}=\f{i}{2\a'} \sqb{\f{k_2^0}{|k_2^0|}- \f{k_1^0}{|k_1^0|}}\times \mc{A}^{I}_{BBB}=\f{i}{2\a'} \sqb{\f{k_2^0}{|k_2^0|}- \f{k_1^0}{|k_1^0|}} \times \mc{A}^{II}_{BBB}
\ee 
For non-trivial physical amplitudes $\sqb{\f{k_2^0}{|k_2^0|}- \f{k_1^0}{|k_1^0|}}=\pm 1$. Thus, upto a sign we recover the 3 point amplitude. This means that generalized amplitudes defined using the mostly BRST exact operator are proportional to the corresponding amplitude verifying our claim. In what follows we shall provide a path integral derivation of the gauge fixing of the two point function and see how mBRST operator arises. 

\section{Origin of the mostly BRST exact operator}

\subsection{The gauge gixing of string amplitudes}
In this section we shall provide the details of how the $V_0$ arises by directly doing the string path integral. We shall follow \cite{Polchinski:1998rq} chapters 3 and 5 in what follows. The S-matrix for  bosonic string theory is given as a weighted sum over all compact 2 dimensional topologies
\be 
S_{j_{1}\cdots j_{n}}\rb{k_1,\cdots,k_n}=\sum_{\substack{\text{compact} \\ \text{topologies}}}\int \f{[dX dg]}{V_{\text{diff}\times \text{Weyl}}}e^{-S_{X}-\l\c}\prod_{i=1}^n\int d^2\s_i \sqrt{g\rb{\s_i}} \mc{V}_{j_i}(k_i,\s_i) \label{S-gen}
\ee
where, $X^m$ are the spacetime coordinates, $g$ denotes the world-sheet metric, $\s^a$ worldsheet coordinates and  $\mc{V}_{j_i}$ are the vertex operators carrying momenta $k_i$ and other quantum labels $j$. The action $S_{X}$ has $\text{diff}\times \text{Weyl}$ redundancies and hence the path integral over all the field configurations $X$ and $g$ is divided by  volume of this group $V_{\text{diff}\times \text{Weyl}}$. Here we are interested in two point amplitudes on a genus 0 surface. Consequently we want to evaluate the following path integral 
\be 
S_{j_{1}, j_{2}}\rb{k_1,k_2}= \int_{g=0} \f{[dX dg]}{V_{\text{diff}\times \text{Weyl}}}e^{-S_{X}-2\l}\prod_{i=1}^2\int d^2\s_i \sqrt{g\rb{\s_i}} \mc{V}_{j_i}(k_i,\s_i) \label{S2}
\ee  
Now consider an infinitesimal diff $\times$ Weyl transformation denoted by $\z$  
\be 
\z : g\rightarrow g^\z\;,\quad g^{\z}_{c d}(\s')=\exp{\rb{2\o(\s)}} \f{\p\s^a}{\p\s'^c}\f{\p\s^b}{\p\s'^b}g_{ab}(\s)
\ee
Let us define the Faddeev-Popov measure $\ud_{FP}$ by 
\be
1=\ud_{FP}(g,\s,X)\int_{diff\times Weyl}[d\z]\d(g-\hat{g}^\z)\prod_{a,(i\in \{1,2\})}\d(\s^a_i-\hat{\s}^{\z a}_i)\d(X^0-\hat{X}^{0\z})    \label{def_FP}
\ee
Let us insert this into \eqref{S2} 
\be 
&&\hspace{-0.4in}S_{j_{1}, j_{2}}\rb{k_1,k_2}\non\\
&=& \int_{g=0} \f{[dX dg]}{V_{\text{diff}\times \text{Weyl}}}e^{-S_{X}-2\l}\prod_{i=1}^2\int d^2\s_i \sqrt{g\rb{\s_i}} \mc{V}_{j_i}(k_i,\s_i)
\non\\
&&\times \ud_{FP}(g,\s,X)\int_{\text{diff}\times \text{Weyl}}[d\z]\d(g-\hat{g}^\z)\prod_{a,(i\in \{1,2\})}\d(\s^a_i-\hat{\s}^{\z a}_i)\;\d(X^0-\hat{X}^{0\z}) \non\\
&=& \int \f{[dX^\z]}{V_{\text{diff}\times \text{Weyl}}}\int[d\z]\;e^{-S_{X^\z}-2\l}\prod_{i=1}^2 \sqrt{\hat{g}^\z\rb{\s_i}} \mc{V}_{j_i}(k_i,\hat{\s}^{\z a}_i)\; \ud_{FP}(\hat{g}^\z,\hat{\s}^\z,X^\z) \d(X^{0\z}-\hat{X}^{0\z})
\ee
where, we integrated over the metric, coordinates $\s_i$ and relabeled the integration variable $X\rightarrow X^\z$. Using the gauge invariance of the delta function, the action, the measure $[dX]$ and the Faddev-Popov determinant and also noting that $\hat{g}^{\z}=\hat{g},\; \hat{\s}^{\z}=\hat{\s}$, we get
\be 
S_{j_{1}, j_{2}}\rb{k_1,k_2}
&=& \int \f{[dX]}{V_{\text{diff}\times \text{Weyl}}}\int[d\z]\;e^{-S_{X}-2\l}\prod_{i=1}^2 \sqrt{\hat{g}\rb{\s_i}} \mc{V}_{j_i}(k_i,\hat{\s}^{a}_i)\; \ud_{FP}(\hat{g},\hat{\s},X) \d(X^{0}-\hat{X}^{0}) \non\\
&=& \int \f{[dX]}{V_{\text{diff}\times \text{Weyl}}}\;e^{-S_{X}-2\l}\prod_{i=1}^2 \sqrt{\hat{g}\rb{\s_i}} \mc{V}_{j_i}(k_i,\hat{\s}^{a}_i)\; \ud_{FP}(\hat{g},\hat{\s},X) \d(X^{0}-\hat{X}^{0}) \times V_{\text{diff}\times \text{Weyl}}\non\\
&=& \int [dX]\;e^{-S_{X}-2\l} \prod_{i=1}^2\sqrt{\hat{g}\rb{\s_i}} \mc{V}_{j_i}(k_i,\hat{\s}^{a}_i)\; \ud_{FP}(\hat{g},\hat{\s},X) \d(X^{0}-\hat{X}^{0}) 
\ee
where in going from the first line to the second line we used the fact that the integrand is independent of $\z$. In order to proceed further we need to calculate $\ud_{FP}$ in the above. This is what we do next. From \eqref{def_FP} we see that 
\be
\ud^{-1}_{FP}(g,\s,X)=\int_{\text{diff}\times \text{Weyl}}[d\z]\d(g-\hat{g}^\z)\prod_{a,(i\in \{1,2\})}\d(\s^a_i-\hat{\s}^{\z a}_i)\d(X^0-\hat{X}^{0\z})   
\ee
We however need, $\ud_{FP}(\hat{g},\hat{\s},X^0)$. But, for $\z$ near identity its action on
the metric $g_{ab}$ and the field $X^{m}$ are given by 
\be 
\d g_{ab}&=&-2(P_1\d\s)_{ab} +(2\d\o-\nabla\cdot \d \s)g_{ab} \non\\
\d X^{m}&=&\p_{a}X^{m}\d\s^a \non
\ee
where, $P_1$ is a differential operator that acts on vectors and produces rank 2, symmetric and traceless tensors, i.e. for $v_a$
\be 
(P_1 v)_{ab}=\f{1}{2}\sqb{\nabla_a v_b+\nabla_b v_a -g_{ab} \nabla_c v^c}
\ee
Consequently, we have 
\be
\ud^{-1}_{FP}(g,\s,X)&=&n_R \int[d\d \o \; d\d\s]\;\d(\d g)\prod_{a,(i\in \{1,2\})}\;\d(\d\s^a(\hat{\s_i}))\;\d(\d X^0)  \non\\
&=&n_R \int[d\d \o \; d\d\s]\;\d\sqb{-\rb{-2(P_1\d\s)_{ab} +(2\d\o-\nabla\cdot \d \s)g_{ab}}}\prod_{(a,i)\in \{1,2\}}\;\d(\d\s^a(\hat{\s_i}))\;\d\sqb{\p_{a}X^{0}\d\s^a}  \non\\
&=&n_R \int[d\d \o \; d\d\s]\;\int [d\b]\exp\left\{2\pi i\int d^2\s \sqrt{\hat{g}}\;\b^{ab}\sqb{2(P_1\d\s)_{ab} -(2\d\o-\nabla\cdot \d \s)g_{ab}}\right\}\non\\
&&\times \int \prod_{a,i=1}^2 dx_{ai}\exp\left\{2\pi i\sum_{(a,i)\in \{1,2\}} x_{ai}\;\d\s^a(\hat{\s_i})\right\}\;\int dy\exp\left\{2\pi i y\; \p_{a}X^{0}\d\s^a\right\}  \non\\
&=&n_R \int[ \; d\d\s] [d\b']\rb{\prod_{a,i=1}^2 dx_{ai}}dy\non\\
&&\times \exp\left\{4\pi i\int d^2\s \sqrt{\hat{g}}\;\b'^{ab}(P_1\d\s)_{ab} +2\pi i\sum_{(a,i)\in \{1,2\}} x_{ai}\;\d\s^a(\hat{\s_i})+2\pi i y \; \p_{a}X^{0}\d\s^a\right\}   
\ee
To get the $\ud_{FP}(g,\s,X)$ we simply replace all the bosonic variables with Grassmann variables i.e. 
\begin{eqnarray}
	\d\s^a &\rightarrow& c^a \\
	\b'^{ab}&\rightarrow& b^{ab} \\
	x_{ai}&\rightarrow& \eta_{ai} \\
	y&\rightarrow& \eta 
\end{eqnarray}
With convenient normalization of the fields write 
\be
\ud_{FP}(\hat{g},\hat{\s},X^0)
&=&n_R \int[ \; dc] [db]\rb{\prod_{a,i=1}^2 d\eta_{ai}}d\eta\non\\
&&\hspace{-0.in}\times \exp\biggl\{4\pi i\int d^2\s \sqrt{\hat{g}}\;b^{ab}(P_1c)_{ab} +2\pi i\sum_{(a,i)\in \{1,2\}} \eta_{ai}\;c^a(\hat{\s_i})+2\pi i \eta \; \p_{a}X^{0}c^a\biggl\}  \non
\ee
Also, with a convenient normalization we define the ghost action from the first term in the exponential above as
\be 
4\pi i\int d^2\s \sqrt{\hat{g}}\;b^{ab}(P_1c)_{ab}\equiv S_{\textup{g}}=\f{1}{2\pi}\int d^2\s \sqrt{\hat{g}}\;b_{ab}\rb{\hat{P}_1c}^{ab}
\ee
Consequently we have 
\be
\ud_{FP}(\hat{g},\hat{\s},X^0)
&=&\f{1}{n_R} \int[ \; dc] [db]\rb{\prod_{a,i=1}^2 d\eta_{ai}}d\eta \; \exp{\rb{-S_{\textup{g}}}}\times  \exp\left\{ 2\pi i\rb{\sum_{(a,i)} \eta_{ai}\;c^a(\hat{\s_i})+\eta \; \p_{a}X^{0}c^a}\right\}   \non\\
&=&\f{1}{n_R}\int[ \; dc] [db]\; \exp{\rb{-S_{\textup{g}}}}\times  \rb{\prod_{(a,i)} \;c^a(\hat{\s_i})} \; \p_{a}X^{0}c^a 
\ee
where we integrated over the variables $\eta, \eta_{ai}$ in order to go the second line. With this we can get the two point function as 
\be 
&&\hspace{-0.4in}S_{j_{1}, j_{2}}\rb{k_1,k_2}\non\\
&=& \f{1}{n_R}\int [dX db\; dc]\;e^{-S_{X}-S_{\textup{g}}-2\l}\prod_{i=1}^2 \sqrt{\hat{g}\rb{\s_i}} \mc{V}_{j_i}(k_i,\hat{\s}^{a}_i)\; \sqb{  \rb{\prod_{(a,i)} \;c^a(\hat{\s_i})} \; \p_{a}X^{0}c^a} \d(X^{0}-\hat{X}^{0}) \non\\\label{S21} 
\ee
This is our final result. Depending on the topology (disk or sphere) and the vertex operators we can extend it to both closed strings and open strings. Let us goto the conformal gauge and where $\hat{g}=e^{2\uo}$. Further in the complex coordinates $z,\bz$ the ghost action becomes \be 
S_{g}=\f{1}{4\pi}\int d^2z \Big(b\bp c+ \tilde{b} \p \tilde{c}\Big)
\ee
and we have have two free field ghost theories  $(b,c)\equiv (b_{zz},c^z)$ and $(\tilde{b},\tilde{c})\equiv (b_{\bz \bz},c^{\bz})$. 
Further analysis depends of whether we are working with open strings or closed strings. We are now going take a look at both of these cases. 
\subsection{Open strings}
In the case of open strings, the set of fields $c,\tilde{c}$ and $b,\tilde{b}$ which live on the complex upper half can be extended to the full complex plane via the doubling trick. Then we only have one set of holomorphic field $(b,c)$. Further the vertex operators are inserted on the boundary and are of the form 
\be 
\int dz_i \sqrt{g(z_i)}\, \mc{V}_{j_i} (z_i) \quad z_i\in \mathbb{R}
\ee
With this understanding the formula \eqref{S21}
for open string becomes 
\be 
&&S_{j_{1}, j_{2}}\rb{k_1,k_2}\non\\
&=& \f{1}{n_R}\int [dX db\; dc]\;e^{-S_{X}-S_{\textup{g}}-2\l} \sqb{\prod_{i=1}^2\sqrt{\hat{g}(\hat{z}_i)} \mc{V}_{j_i}(k_i,\hat{\z}_i)c(\hat{z}_i)} \, c(z_0)\p X^0\, \d(X^{0}-\hat{X}^{0})\non\\
&\propto &\Big\la c(\hat{z}_1)\mc{V}_{j_1}(k_1,\hat{z}_1) \; c(\hat{z}_1)\mc{V}_{j_2}(k_2,\hat{z}_2)\;c(z_0)\p X^0\, \d(X^{0}-\hat{X}^{0}) \Big\ra
\ee
On a disk $c$ ghost has three zero modes. Consequently the above amplitude is non-zero. The $c$ ghost coming from fixing the $X^0$ comes with a factor of $\p X^0$ 
\be 
c(z_0)\p X^0\, \d(X^{0}-\hat{X}^{0})                                             \label{mBRST1}
\ee 
This piece will form a part of the mBRST exact operator. 
\subsection{Closed strings}  \label{closed_strings}
 For closed strings the $c,\tilde{c}$ and $b,\tilde{b}$ remain independent and are complex conjugates of each-other. We denote this by $\tilde{c}=\bc$ and $\tilde{b}=\bar{b}$. With this its easy to write the square bracket in \eqref{S21}   
\be 
  \rb{\prod_{(a,i)} \;c^a(\hat{\s_i})} \; \p_{a}X^{0}c^a&=&c(z_1)\bc(\bz_1)\,c(z_2)\bc(\bz_2)\rb{c\p X^0+\bc \bp X^0} \non\\
  &=& c(z_1)\bc(\bz_1)\,c(z_2)\bc(\bz_2) c\p X^0 +c(z_1)\bc(\bz_1)\,c(z_2)\bc(\bz_2)\bc \bp X^0\label{2ptcghost}
\ee
The first term fails to saturate the $\bc$ zero modes, while the second term fails to saturate the $c$ zero modes. Consequently the amplitude vanishes. The root cause of the problem is that we didn't fix the conformal group completely. In the case of sphere, there are 6 real CKVs. We implicitly fixed 5 of them when we wrote the ansatz for the $\ud_{FP}$ in \eqref{def_FP}. Let us see how. Fixing the position of the two closed strings fixes 4 of the CKVs, while fixing $X^0$ fixes only one CKV as $\d X^0 =\d\s^a \p_a X^0=\d\s^1 \p_1 X^0+\d\s^2 \p_2 X^0$. The remedy perhaps is to fix another spacetime coordinate, say $X^9$ as well. This will fix all of the CKVs on the sphere. Whether it leads to something sensible is a matter of further investigation, which we leave for future work.

\section{Introducing the mBRST exact operator} \label{mBRSTintro}
We shall now add another piece to \eqref{mBRST1} that gives vanishing contribution to the amplitude when the external momenta satisfy the $\sum_{i}k_i=0$. This will finally give us the mBRST exact operator. We reiterate that our interest behind do so is because this relation can be readily lifted to the pure spinor formalism. Consider the following OPE between the BRST current $j_B(z)$ and a matter field operator $\mc{O}^m(0)$ in bosonic string theory \cite{Polchinski:1998rq}

$$j_{B}(z)\mc{O}^m(0)\sim \f{h}{z^2}c\,\mc{O}^m(0)+\f{1}{z}\sqb{h\rb{\p c}\mc{O}^m(0)+c\p O^{m}(0)}$$
Hence, we will have 

$$
\oint dz j_{B}(z)O^m(0)=h\rb{\p c}\mc{O}^m(0)+c\p O^{m}(0)
$$
or in other words 
\be
[Q_B,O^{m}(z)]=h\rb{\p c}\mc{O}^m(z)+c\p O^{m}(z)
\ee
For an operator of the form $:e^{ik\cdot X}:$ we have 
\be
[Q_B,:e^{ik\cdot X(z)}:]=h\rb{\p c}(z):e^{ik\cdot X(z)}: + i \rb{k\cdot \p X} c\, :e^{ik\cdot X(z)}:
\ee
which has coformal dimension $\a' k^2$. Our interest is in  $k_0^m=q t^m$, $q$ is a time-like unit vector so that $t^2=-1$ and q is a real parameter so that $h=-\a' q^2$ and hence 
\be
[Q_B,:e^{ik_0\cdot X(z)}:]&=&-\a' q^2\rb{\p c}(z):e^{ik_0\cdot X(z)}: +i \rb{k\cdot \p X} c\, :e^{k_0\cdot X(z)}: \non\\
&=&-\a' q^2\rb{\p c}(z):e^{-iq X^0(z)}: - iq\,\p X^0 \,c\, :e^{-iq X^0(z)}: 
\ee
where, in going to the second line we made a choice of $t_m=(-1,0,\cdots,0)$. Consequently, 
\be 
\int_{-\infty}^{\infty} \f{dq}{q} \;[Q_B,:e^{ik_0\cdot X(z)}:]&=&-\a'\rb{\p c}(z)\int_{-\infty}^{\infty} dq \; q\; :e^{-iq X^0(z)}:\quad-i\p X^0 (z)\int_{-\infty}^{\infty} dq\; :e^{-iq X^0(z)}:\non\\
&=&-\a' \rb{\p c}(z)\int_{-\infty}^{\infty} dq \; \f{ \p }{-i\p X^0}\rb{e^{-iqX^0}}+2\pi \; c(z)\;\p X^0 (z)  \d(X^{0}(z))\non\\
&=&\a'\rb{\p c}(z)\f{ \p }{i\p X^0}\int_{-\infty}^{\infty} dq \; e^{-iqX^0}+2\pi \; c(z)\;\p X^0 (z)  \d(X^{0}(z))\non\\
&=&2\pi \sqb{\a'\rb{\p c}(z)\f{ \p}{\p X^0}\d(X^{0}(z))+\; c(z)\;\p X^0 (z)  \d(X^{0}(z))}
\ee
Let us define 
\be 
V_{0}\equiv \int_{-\infty}^{\infty} \f{dq}{2\pi  q} \;[Q_B,:e^{-iqX^0(z)}:]=c(z)\;\p X^0 (z)  \d(X^{0}(z))+\a'\,\p c(z)\; \d'\rb{X^{0}(z)} \label{V0}
\ee

We saw in the previous subsection that the first term in \eqref{V0} naturally appears via a Faddeev Popov gauge fixing of the string two point amplitude for the open strings. The second term is something that has a special feature that it vanishes when the momenta of the external strings sum to zero. In order to see this, we note that it comes from $-\a'\rb{\p c}\int_{-\infty}^{\infty} dq \; q\; e^{-iq X^0}(z)$, which has following action in a correlation function
\be 
\la -\a'\rb{\p c}\int_{-\infty}^{\infty} dq \; q\; e^{-iq X^0}(z)\cdots \ra=\int dq \, q \,\d\rb{-q+\sum_{i=1}^{n}k_i^0}\;\d\rb{\sum_{i=1}^{n}\vec{k}_i}\cdots
\ee 
If our interest lies in the amplitudes satisfying $\d\rb{\sum_{i=1}^nk_i}$, the above term vanishes. Thus, we shall insert the mBRST exact operator into the amplitude with only extra condition that the external momenta add upto zero. In the pure spinor case we have 
\be 
\sqb{Q,e^{-iqX^0}}=-\f{iq\a'}{2}\rb{\l\g^0\t}e^{-iqX^0}
\ee
Thus, the mBRST exact operator in the pure spinor  formalism takes the form 
\be 
 V_{0}\equiv \int_{-\infty}^{\infty} \f{dq}{2\pi  q} \;[Q_B,:e^{-iqX^0(z)}:]=\f{\a'}{4\pi i}\int_{-\infty}^{\infty}dq \rb{\l\g^0\t}:e^{-iqX^0(z)}:
\ee
Further, we can covariantize the above operator by making use of a timelike vector $t$
\be 
V_0(z)=\f{\a'}{4\pi i}\int_{-\infty}^{\infty}dq\; t_m\rb{\l\g^m\t}:e^{-iq t_n X^n(z)}: \, ,  \qquad t^2=-1
\ee
\section{Discussion}  \label{discuss}
We have seen that the utility of mBRST exact operator in fixing conformal killing vectors in tree level amplitudes goes beyond just two point amplitudes, though it is in this context that it appeared. We recall that conformal killing vectors appear at one loop amplitudes as well. It is a natural question is if this operator can be used to fix the CKV that appear at the one loop, which we wish to explore in future work.

Another utility of this operator is that it can be used to determine the normalization of the vertex operator. $V_0$ is defined upto a constant. This constant can be fixed by calculating a given amplitude with and without use of the mBRST exact operator e.g. by comparing the three point amplitude computed in two ways we did in the main text. After this constant has been fixed, we can calculate the two point function with the vertex operator in concern. Then we can determine the normalization of the vertex operator by demanding a unitarity requirement of two point amplitudes \cite{Erbin:2019uiz} 
\be 
\mc{A}_{2}(\vec{k_1},\vec{k}_2)=\int \f{d^{D-1}}{(2\pi)^{D-1}}\f{1}{2k^0}\mc{A}_{2}(\vec{k_1},\vec{k}_2)\, \mc{A}_{2}(\vec{k_1},\vec{k}_2)
\ee
This may significantly reduce the effort as otherwise we need to rely on factorization of higher point amplitudes. 

 We saw that there is an issue for closed strings. A possible resolution of this is to fix any one of the space coordinates $X^i\rightarrow \hat{X}^{i\z}$ say $X^9$, just like the time coordinate as we have done in this paper.  This procedure will produce 
\be 
\rb{c^a(\s)\p_a X^0(\s)\; \d(X^{0}-\hat{X}^{0})}\rb{c^a(\s)\p_a X^9(\s)\; \d(X^{9}-\hat{X}^{9})}
\ee
which can be repackaged into a product of two mBRST exact operators with parameters $q_1$ and $q_2$ respectively. This should work so long as the external momenta keep satisfying $\sum_{i}k_i=0$ (just as in this work). We shall have two real integrals over $q_1$ and $q_2$ and the physical amplitudes will be obtained by extracting the $q_1\rightarrow 0, q_2\rightarrow 0$ contribution. This also resonates with the holomorphic factorization. 
\vspace{.25in}

\noindent{\bf Acknowledgments: } We thank Mritunjay Verma for useful discussions and comments on the draft.  We also thank the people and the Government of India for their continuous support for carrying out research in theoretical physics.

\appendix
\section{Pure spinor formalism important formulae} \label{ps_results}
In this appendix we recall some of the results from the PS formalism literature that we used in main text. 

\subsection{List of OPEs}
\be
\Pi^m(z)V(w)&\simeq& -\f{\a'}{z-w} \p^m V \label{piOPE}\\
d_{\a}(z) V(w)&\simeq& \f{\a'}{2(z-w)}D_\a \label{dOPE}V\\
N^{mn}(z) \l^\a(w) &\simeq& \f{\a' \rb{\g^{m n}}^\a_{\;\;\b} \l^\b(w)}  {4\; (z-w)}\label{NOPE}
\ee
where, $V$ is any function of $\l^\a,\t^\a,X^m$. In the above $D_\a$ is the super-Poincare covariant derivative on superspace 
\be 
D_\a =\f{\p}{\p\t^\a}+\rb{\g^m\t}_{\a}\f{\p}{\p X^m}
\ee
\subsection{Pure spinor superspace identities}
We only require a couple of PSS identities in this work which are stated below
\be 
\la\rb{\l\g^m\t} \rb{\l\g^n\t} \rb{\l\g^p\t}\rb{\t\g_{s t u}\t}\ra=\f{1}{120}\delta^{m n p}_{s t u}  \label{id1}
\ee

\be 
\la\rb{\l\g_m\t} \rb{\l \g_n\t} \rb{\l \g^{pqr}\t}\rb{\t\g_{s t u}\t}\ra=\f{1}{70}\d^{[p}_{[m}\d_{n][s}\d^{q}_{t}\d^{r]}_{u]}  \label{id2}
\ee
where, $\d^{a_1a_2\cdots a_n}_{b_1 b_2\cdots b_n}\equiv \d^{a_1}_{[b_1}\d^{a_2}_{b2}\cdots \d^{a_n}_{b_n]}$. For a more complete list see \cite{Berkovits_mafra_non_minimal}.
\subsection{Theta expansions}
 We shall be requiring the theta expansion of all of superfields  appearing in 10 dimensional sYM. We refer the reader to the appendix of \cite{Chakrabarti:2018bah} for a detailed derivation. Here, we shall recall the relevant results. The main result that we require is the recursive relations among the components of the various superfields i.e. 
\be
A^{(\ell)}_\alpha &=& \f{2}{1+\ell}A^{(\ell-1)}_m (\gamma^m\theta)_\alpha
\non\\[.3cm]
\quad  A^{(\ell)}_m &=& -\f{1}{\ell}(\theta\gamma_m W^{(\ell-1)})  \non\\[.3cm]
W^\alpha_{(\ell)} &=&\f{1}{2\ell}(\gamma^{mn}\theta)^\alpha\ F^{(\ell-1)}_{mn}\non\\[.3cm]
F^{(\ell)}_{mn}& =&\p_mA^{(\ell)}_n-\p_nA^{(\ell)}_m
\ee
To start the recursion we need the lowest component of the superfields $A^{(0)}_{m}=a_m$ and $W_{(0)}^\a=\c^\a$. With this we find 
\be
A_\alpha &=& a_m(\gamma^m\theta)_\alpha -\f{2}{3}(\gamma^m\theta)_\alpha (\theta\gamma_m\chi)-\f{1}{8}(\gamma_m\theta)_\alpha (\theta\gamma^{mpq}\theta)f_{pq}-\f{i}{15}(\gamma_m\theta)_\alpha (\theta\gamma_p\chi)(\theta\gamma^{mpq}\theta)k_q+\cdots\non\\[.4cm]
A_m&=& a_m -(\theta\gamma_m\chi) -\f{1}{4}(\theta\gamma_{mnp}\theta)f^{np}\ -\ \f{1}{6}(\theta\gamma_{m}\gamma^{pq}\theta)(\theta\gamma_{[m}\p_{n]}\chi) +\f{1}{48}  (\theta\gamma_{m}\gamma^{rn}\theta)(\theta\gamma_{npq}\theta)\p_rf^{pq}+\cdots\non\\[.4cm]
W^\a&=&\c^\alpha + \p_m a_{n} \;\rb{\g^{m n} \t}^\a  
-\f{1}{2}\rb{\t\g_{n} \p_m\c} \rb{\g^{m n} \t}^\a	-\f{1}{6} \p_m \p_s a_{t}\rb{\t\g_{n}\g^{s t} \t} \rb{\g^{m n} \t}^\a +\cdots 
\ee
where, $f_{mn}\equiv\p_{[m}a_{n]} \equiv \p_m a_n -\p_n a_m$. Further we expand the gluon and the gluino fields in plane wave 
\be 
a_m=\e_m e^{ik\cdot X}, \quad \c^\a=\x^\a e^{ik\cdot X}
\ee 
\section{Three gluon amplitude using standard prescription} \label{3pt}
In this appendix we provide the details of the standard computation of 3 point amplitude in open strings in the pure spinor formalism.
For the direct computation of the three point amplitude we only require the superfield $A_{\a}$. Let us calculate the following
\be 
\la V_1(z_1) V_2(z_2) V_3(z_3)\ra =\la \l^\a A^1_\a \, \l^\b A^2_\b \, \l^\g A^3_\g\ra\times \la :e^{ik\cdot X(z_1)}: \, :e^{ik\cdot X(z_2)}:\, :e^{ik\cdot X(z_3)}:\ra
\ee
On making use of 
\be 
\left\la \prod_{i=1}^{n} :e^{ik\cdot X(z_i)}:\right\ra=i C^X_{D_2}(2\pi)^d\d(\sum_{i}k_i)\prod_{i<j}|z_i-z_j|^{2\a'k_i\cdot k_j}
\ee
we find that 
\be 
\la :e^{ik\cdot X(z_1)}: \, :e^{ik\cdot X(z_2)}:\, :e^{ik\cdot X(z_3)}:\ra=i C^X_{D_2}(2\pi)^d\d(\sum_{i}k_i) |z_1-z_2|^{2\a'k_1\cdot k_2} \,|z_1-z_3|^{2\a'k_1\cdot k_2}\,|z_2-z_3|^{2\a'k_2\cdot k_3}\non
\ee
On choosing $z_1=0,z_2=1,z_3=\inf$ and using momentum conservation along with $k_i^2=0$ we find that all coordinate dependence goes away. 
Hence, we have
\be
\la :e^{ik\cdot X(z_1)}: \, :e^{ik\cdot X(z_2)}:\, :e^{ik\cdot X(z_3)}:\ra=i C^X_{D_2}(2\pi)^d\d(\sum_{i=1}^3k_i)\equiv \ug_3                      \label{g3}
\ee
We can evaluate the $(\l^\a,\t^\b)$ correlator by making use of the following PSS identity in \eqref{id1} 
\be 
\mc{M}^{I}_{BBB}&\equiv &\la \l^\a A^1_\a \, \l^\b A^2_\b \, \l^\g A^3_\g\ra\non\\
&=&\f{i }{360}\Big[ \rb{\e_1\cdot \e_2}\rb{\e_3\cdot k_{12}}+\rb{\e_1\cdot \e_3}\rb{\e_2\cdot k_{31}}+\rb{\e_2\cdot \e_3}\rb{\e_1\cdot k_{23}}\Big] \non\\
&=&\f{i }{180}\Big[ \rb{\e_1\cdot \e_2}\rb{\e_3\cdot k_{1}}+\rb{\e_1\cdot \e_3}\rb{\e_2\cdot k_{3}}+\rb{\e_2\cdot \e_3}\rb{\e_1\cdot k_{2}}\Big]
\ee
where, $k_{ij}\equiv k_i-k_j$ and we goto the second line by making use of the momentum conservation $\sum_i k_i=0$ and $\e_i\cdot k_i=0$. Further, the superscript $I$ on $\mc{M}$ indicates that its the ordering of the vertex operators as $(1,2,3)$. 

Consequently 
\be 
\mc{A}^{I}_{BBB}=\f{i }{180}\Big[ \rb{\e_1\cdot \e_2}\rb{\e_3\cdot k_{1}}+\rb{\e_1\cdot \e_3}\rb{\e_2\cdot k_{3}}+\rb{\e_2\cdot \e_3}\rb{\e_1\cdot k_{2}}\Big]\ug_{3}=\mc{M}^{I}_{BBB}\ug_3
\ee

\section{Three point amplitude using mBRST exact operator} \label{3pt_BRST exact}
In this appendix we calculate the sum of $T_2,T_3,T_4$ in \eqref{T2}-\eqref{T4}that appear in the computation of $\mc{S}_4$. $T_1$ vanishes identically since there are no $p_\a$ in the correlator to eliminate the $\p\t^\a$. $T_2,T_3$ and $T_4$ each can be computed in terms of the superfields and simplified to 
\be 
&&T_1+T_2+T_3+T_4\non\\
&=&\f{\a'}{z_{03}}\biggl[i k_0^m t_s\left\la  \rb{\l\g^s\t} \; V_1\; V_2\; A^{\rb{3}}_{m} \right\ra - \f{t_s}{2} \left\la V_1(z_1) V_2(z_2) \rb{\l\g^s W^{(3)}}\right\ra \non\\
&&+\f{t_n}{2} \Big\la (\l \g^m \t)\; V_1 \; V_2 \; i\rb{k_{3m} A^{(3)}_{n}-k_{3n} A^{(3)}_{m}}\Big\ra +\f{t_s}{2} \Big\la (\l\g^{m n s} \t)\; V_1 \; V_2 \; ik_{3m}A^{(3)}_{n}  \Big\ra\biggl] \non\\
&&+\f{\a't_s}{z_{13}} \Big[i k_1^m \left\la  \rb{\l\g^s\t} \; V_1\; V_2\; A^{\rb{3}}_{m} \right\ra+\left\la \rb{\l\g^s\t} A^{(1)}_m  V_2(z_2) \rb{\l\g^m W^{(3)}} \; \right\ra\Big]\non\\
&&+\f{\a'}{z_{23}} \Big[i k_2^m t_s\left\la  \rb{\l\g^s\t} \; V_1\; V_2\; A^{\rb{3}}_{m} \right\ra - \left\la V_0 \;   V_1(z_1) \;A^{(2)}_m\; \rb{\l\g^m W^{(3)}} \; \right\ra\Big] 
\ee
We can calculate each of the correlator by making use of the pure spinor superspace identities \eqref{id1} and \eqref{id2}. The terms corresponding to $z_{03}$  are given by  
\be 
&&-\f{1}{288} \rb{\e_3\cdot t} \rb{\e_1\cdot k_2} \rb{\e_2\cdot k_3} -\f{1}{720} \rb{\e_3\cdot t} \rb{\e_1\cdot \e_2} s +\f{1}{288} \rb{\e_3\cdot t} \rb{\e_1\cdot k_3} \rb{\e_2\cdot k_1}
 \non\\
 &&+\f{1}{720} \rb{\e_3\cdot t} \rb{\e_1\cdot \e_2} t +\f{1}{288} \rb{k_3\cdot t} \rb{\e_1\cdot k_2} \rb{\e_2\cdot \e_3}-\f{1}{360} \rb{k_3\cdot t} \rb{\e_1\cdot \e_2} \rb{\e_3\cdot k_2} \non\\
 && -\f{1}{288} \rb{k_3\cdot t} \rb{\e_1\cdot \e_3} \rb{\e_2\cdot k_1}+\f{1}{360} \rb{k_3\cdot t} \rb{\e_1\cdot \e_2} \rb{\e_3 \cdot k_1} -\f{1}{480} \rb{k_3\cdot t} \rb{\e_1\cdot k_3} \rb{\e_2\cdot \e_3}\non\\
 &&+\f{1}{480} \rb{k_3\cdot t} \rb{\e_1\cdot \e_3} \rb{\e_2\cdot k_3}+\f{1}{1440} \rb{k_0\cdot t} \rb{\e_1\cdot k_3} \rb{\e_2\cdot \e_3}-\f{1}{1440} \rb{k_0\cdot t} \rb{\e_1\cdot \e_3} \rb{\e_2\cdot k_3} \non\\
 &&+\f{1}{360} \rb{\e_2\cdot t} \rb{\e_1\cdot k_2} \rb{\e_3\cdot k_1}+\f{1}{360} \rb{\e_2\cdot t} \rb{\e_1\cdot k_2} \rb{\e_3\cdot k_2}-\f{1}{360} \rb{k_2\cdot t} \rb{\e_1\cdot \e_2} \rb{\e_3\cdot k_1}\non\\
 &&-\f{1}{360} \rb{k_2\cdot t} \rb{\e_1\cdot \e_2} \rb{\e_3\cdot k_2}-\f{1}{360} \rb{\e_1\cdot t} \rb{\e_1\cdot k_1} \rb{\e_3\cdot k_1}
 -\f{1}{360} \rb{\e_1\cdot t} \rb{\e_2\cdot k_1} \rb{\e_3\cdot k_2}\non\\
 &&+\f{1}{360} \rb{k_1\cdot t} \rb{\e_1\cdot \e_2} \rb{\e_3\cdot k_1}+\f{1}{360} \rb{k_1\cdot t} \rb{\e_1\cdot \e_2} \rb{\e_3\cdot k_2} \label{z03}
\ee
 In order to simplify this further we recall that we only consider terms where the $\sum_{i=1}^{3}k_i=0$. In this case all the terms that are proportional to Mandelstam variables will vanish identically. This is because they all are proportional to $q$ (see \eqref{mandelstam1}) on account of the overall $\d\rb{q-\sum_{i=1}^3k_i}$ in $\mc{S}_4$.
 Further the terms that do not involve contraction among polarization tensors add upto zero 
 \be 
 &&-\f{1}{288} \rb{\e_3\cdot t} \rb{\e_1\cdot k_2} \rb{\e_2\cdot k_3}  +\f{1}{288} \rb{\e_3\cdot t} \rb{\e_1\cdot k_3} \rb{\e_2\cdot k_1}
 \non\\
 &&+\f{1}{360} \rb{\e_2\cdot t} \rb{\e_1\cdot k_2} \rb{\e_3\cdot k_1}+\f{1}{360} \rb{\e_2\cdot t} \rb{\e_1\cdot k_2} \rb{\e_3\cdot k_2}\non\\
 &&-\f{1}{360} \rb{\e_1\cdot t} \rb{\e_1\cdot k_1} \rb{\e_3\cdot k_1}
 -\f{1}{360} \rb{\e_1\cdot t} \rb{\e_2\cdot k_1} \rb{\e_3\cdot k_2} \label{z03a}
 \ee
Each of the lines in the above equation adds to zero. Notice e.g. 
$$ 
\rb{e_1\cdot k_3}=-\rb{e_1\cdot k_2} -\rb{e_1\cdot k_0}=-\rb{e_1\cdot k_2} -q\rb{e_1\cdot t}
$$
the second term being proportional to $q$ will vanish after $q$ integration. With this we can see that the second term in first line is precisely the first term with opposite sign. The second and third line vanish by noticing that on momentum conservation we will produce a term proportional to q and another term that vanishes due to polarization condition. Using similar manipulations we finally find that rest of the terms  add to give 
\be 
\f{\rb{k_3\cdot t} }{360}\Big[ \rb{\e_1\cdot \e_2}\rb{\e_3\cdot k_{12}}+\rb{\e_1\cdot \e_3}\rb{\e_2\cdot k_{31}}+\rb{\e_2\cdot \e_3}\rb{\e_1\cdot k_{23}}\Big]
\ee
where$k_{ij}\equiv k_i-k_j$. Terms corresponding to $z_{13}$ are given by 
\be 
&&\f{1}{360} \rb{\e_2\cdot t} \rb{\e_1\cdot k_3} \rb{\e_3\cdot k_2} -\f{1}{360} \rb{k_2\cdot t} \rb{\e_1\cdot k_3} \rb{\e_2\cdot \e_3} -\f{1}{1440} \rb{\e_1\cdot t} \rb{\e_2\cdot k_3} \rb{\e_3\cdot k_2}
\non\\
&&-\f{1}{2880} \rb{\e_1\cdot t} \rb{\e_2\cdot \e_3} s +\f{1}{720} \rb{\e_2\cdot t} \rb{\e_1\cdot \e_3} s +\f{1}{360} \rb{k_2\cdot t} \rb{\e_1\cdot \e_3} \rb{\e_2\cdot k_3}\non\\
&& +\f{1}{1440} \rb{\e_3\cdot t} \rb{\e_1\cdot k_2} \rb{\e_2\cdot k_3}+\f{1}{2880} \rb{\e_3\cdot t} \rb{\e_1\cdot \e_2} s -\f{1}{1440} \rb{k_3\cdot t} \rb{\e_1\cdot k_2} \rb{\e_2\cdot \e_3}\non\\
&&+\f{1}{1440} \rb{k_3\cdot t} \rb{\e_1\cdot \e_2} \rb{\e_3\cdot k_2}+\f{1}{960} \rb{\e_1\cdot t} \rb{\e_2\cdot \e_3} t + \f{1}{360} \rb{k_1\cdot t} \rb{\e_1\cdot k_3} \rb{\e_2\cdot \e_3} \non\\
&&-\f{1}{360} \rb{\e_3\cdot t} \rb{\e_1\cdot k_3} \rb{\e_2\cdot k_1}-\f{1}{960} \rb{\e_3 \cdot t} \rb{\e_1\cdot \e_2} t+\f{1}{480} \rb{\e_1\cdot t} \rb{\e_2\cdot k_3} \rb{\e_3\cdot k_1}\non\\
&&-\f{1}{360} \rb{k_1\cdot t} \rb{\e_1\cdot \e_3} \rb{\e_2\cdot k_3}+\f{1}{360} \rb{k_3\cdot t} \rb{\e_1\cdot \e_3} \rb{\e_2\cdot k_1}
-\f{1}{480} \rb{k_3\cdot t} \rb{\e_1\cdot \e_2} \rb{\e_3\cdot k_1}\non\\
&&-\f{1}{480} \rb{\e_3\cdot t} \rb{\e_1\cdot k_3} \rb{\e_2\cdot k_3}+\f{1}{480} \rb{k_3\cdot t} \rb{\e_1\cdot k_3} \rb{\e_2\cdot \e_3} -\f{1}{360} \rb{e_2\cdot t} \rb{\e_1\cdot k_2} \rb{\e_3\cdot k_1}\non\\
&&+\f{1}{360} \rb{k_2\cdot t} \rb{\e_1\cdot \e_2} \rb{\e_3\cdot k_1}+\f{1}{360} \rb{\e_1\cdot t} \rb{\e_2\cdot k_1} \rb{\e_3\cdot k_1}-\f{1}{360} \rb{k_1\cdot t} \rb{\e_1\cdot \e_2} \rb{\e_3\cdot k_1}                                                     \label{z13}
\ee
These can be simplified to 
\be 
\f{\rb{k_2\cdot t} }{360}\Big[ \rb{\e_1\cdot \e_2}\rb{\e_3\cdot k_{12}}+\rb{\e_1\cdot \e_3}\rb{\e_2\cdot k_{31}}+\rb{\e_2\cdot \e_3}\rb{\e_1\cdot k_{23}}\Big]
\ee
Similarly we can evaluate $z_{23}$, but, since $z_2=\infty$ the coefficient of this terms go to zero. Let us mention the result for completeness
\be
&&-\f{1}{360} \rb{\e_1\cdot t} \rb{\e_2\cdot k_3} \rb{\e_3\cdot k_1} +\f{1}{360} \rb{k_1\cdot t} \rb{\e_1\cdot \e_3} \rb{\e_2\cdot k_3} +\f{1}{1440} \rb{\e_2\cdot t} \rb{\e_1\cdot k_3} \rb{\e_3\cdot k_1}
\non\\
&&+\f{1}{2880} \rb{\e_2\cdot t} \rb{\e_1\cdot \e_3} t -\f{1}{720} \rb{\e_1\cdot t} \rb{\e_2\cdot \e_3} t -\f{1}{360} \rb{k_1\cdot t} \rb{\e_1\cdot k_3} \rb{\e_2\cdot \e_3}\non\\
&& -\f{1}{1440} \rb{\e_3\cdot t} \rb{\e_1\cdot k_3} \rb{\e_2\cdot k_1}-\f{1}{2880} \rb{\e_3\cdot t} \rb{\e_1\cdot \e_2} t +\f{1}{1440} \rb{k_3\cdot t} \rb{\e_1\cdot \e_3} \rb{\e_2\cdot k_1}\non\\
&&-\f{1}{1440} \rb{k_3\cdot t} \rb{\e_1\cdot \e_2} \rb{\e_3\cdot k_1}-\f{1}{960} \rb{\e_2\cdot t} \rb{\e_1\cdot \e_3} s - \f{1}{360} \rb{k_2\cdot t} \rb{\e_1\cdot \e_3} \rb{\e_2\cdot k_3} \non\\
&&+\f{1}{360} \rb{\e_3\cdot t} \rb{\e_1\cdot k_2} \rb{\e_2\cdot k_3}+\f{1}{960} \rb{\e_3 \cdot t} \rb{\e_1\cdot \e_2} s -\f{1}{480} \rb{\e_2\cdot t} \rb{\e_1\cdot k_3} \rb{\e_3\cdot k_2}\non\\
&&+\f{1}{360} \rb{k_2\cdot t} \rb{\e_1\cdot k_3} \rb{\e_2\cdot \e_3}-\f{1}{360} \rb{k_3\cdot t} \rb{\e_1\cdot k_2} \rb{\e_2\cdot \e_3}
+\f{1}{480} \rb{k_3\cdot t} \rb{\e_1\cdot \e_2} \rb{\e_3\cdot k_2}\non\\
&&+\f{1}{480} \rb{\e_3\cdot t} \rb{\e_1\cdot k_3} \rb{\e_2\cdot k_3}-\f{1}{480} \rb{k_3\cdot t} \rb{\e_1\cdot \e_3} \rb{\e_2\cdot k_3} -\f{1}{360} \rb{e_2\cdot t} \rb{\e_1\cdot k_2} \rb{\e_3\cdot k_2}\non\\
&&+\f{1}{360} \rb{k_2\cdot t} \rb{\e_1\cdot \e_2} \rb{\e_3\cdot k_2}+\f{1}{360} \rb{\e_1\cdot t} \rb{\e_2\cdot k_1} \rb{\e_3\cdot k_2}-\f{1}{360} \rb{k_1\cdot t} \rb{\e_1\cdot \e_2} \rb{\e_3\cdot k_2}                                                     \label{z23}
\ee
which simplifies to 
\be 
\f{\rb{k_1\cdot t} }{360}\Big[ \rb{\e_1\cdot \e_2}\rb{\e_3\cdot k_{12}}+\rb{\e_1\cdot \e_3}\rb{\e_2\cdot k_{31}}+\rb{\e_2\cdot \e_3}\rb{\e_1\cdot k_{23}}\Big]
\ee

\bibliographystyle{JHEP}
\bibliography{reference}
\end{document}